\documentclass[aps,pre,twocolumn,groupedaddress]{revtex4}
\usepackage{graphicx}
\usepackage{amssymb}
\usepackage{color}
\newcommand{ \be}{\begin{equation}}
\newcommand{ \ee}{\end{equation}}
\begin{document}

\title{Critical fluctuations and anomalous transport in soft Yukawa-Langevin systems}

\author{S. Ratynskaia,$^{1}$ G. Regnoli,$^{2,1}$ K. Rypdal,$^{3}$ B. Klumov,$^{2}$  G. Morfill$^{2}$}
\date{\today}
\affiliation{$^1$Space-and Plasma Physics, Royal Institute of Technology, Stockholm, Sweden
\\$^2$Max-Planck-Institut f\"ur extraterrestrische
Physik, Garching, Germany
\\$^3$Department of Physics and Technology, University of Troms{\o}, Norway}

\begin{abstract}
Simulation of  a Langevin-dynamics model demonstrates emergence of critical fluctuations and anomalous grain transport which have been   observed in experiments on ``soft'' quasi-two-dimensional dusty plasma clusters. Our model does not contain external drive or plasma interactions that serve to drive the system away from thermodynamic equilibrium. The grains  are confined by an external potential,  interact via static Yukawa forces, and are subject to  stochastic heating and dissipation from neutrals. One remarkable feature is emergence of leptokurtic probability distributions of grain displacements $\xi(\tau)$ on time-scales $\tau<\tau_{\Delta}$ , where $\tau_{\Delta}$ is the time at which the standard deviation $\sigma(\tau)\equiv \langle \xi^2(\tau) \rangle^{1/2}$ approaches  the mean inter-grain distance $\Delta$. Others are development of humps in the distributions on multiples of $\Delta$, anomalous Hurst exponents, and transitions from leptokurtic towards Gaussian displacement distributions on time scales $\tau>\tau_{\Delta}$. The latter is a signature of intermittency, here interpreted as a transition from bursty  transport  associated with hopping on intermediate time scales to vortical flows on longer time scales. These intermittency features are quantitatively modeled by a single-partice It{\^o}-Langevin stochastic equation with a non-linear drift term.
\end{abstract}

\maketitle

\section{Introduction}
Since a large aggregate of dust grains embedded in a
plasma can attain virtually all known states of matter \cite{review}, the ability
of  optical tracking of individual  grain trajectories has made these systems particularly attractive for studying  transport  properties and
emergent  complex behavior in soft materials.

In a number of recent dusty plasmas experiments anomalous properties, like non-Gaussian displacement distributions and anomalous diffusion, have been observed.
Yet it is not clear whether this behavior  derives from non-equilibrium physics due to external forcing of the grain aggregate, or whether
critical fluctuations \cite{Christensen} and anomalous transport may emerge in aggregates subject exclusively to statically screened Coulomb
 forces and independent stochastic forcing and friction due to collisions with the background neutral gas. The purpose of this paper is to
  demonstrate that the latter is possible, and that non-Gaussian statistics does not have to arise from particular
   non-equilibrium physics.

So far few molecular-dynamics (MD) or Langevin-dynamics (LD) simulations of dust transport have been able to reproduce the anomalous results observed in experiments, although subdiffusion on short time scales have recently been reported for a bidispersive disordered assembley of Yukawa particles \cite{Reichardt2007}. Here we adopt the terminology employed in \cite{GoreePRE2008}, where MD refers to simulation of an aggregate of particles interacting via Yukawa forces and LD refers to a system that in addition is subject to stochastic forcing and friction represented by a Langevin thermostat. The latter represents a heat bath which should allow the system to relax to thermodynamic equilibrium with the neutral gas.

Under ground-based laboratory conditions dust grains levitate in electrostatic sheaths,  which give rise to  plasma anisotropy and supra-thermal ion flows that may act as external forcing. Such experiments  are referred to as two-dimensional (2D or quasi-2D) due to the fact that vertical motion of particles is severely restricted (monolayers) and/or particles are vertically aligned due to the ion flow.
Dust transport under such conditions has been extensively studied both in laboratory \cite{LinI1998,LinI2001,LinI2002,LinI2004,LinIPPCF2004,LinIPPCF2005,PoP,PRL,GoreePRE2007,GoreePRL2008,GoreePRE2008,NPJ,Knapek} and   simulations
 \cite{GoreePRE2007,GoreePRL2008,GoreePRE2008,NPJ,Knapek}. Anomalous transport exhibiting  Gaussian \cite{GoreePRE2007,GoreePRE2008} as well as non-Gaussian \cite{LinI2002,LinIPPCF2004,PoP,PRL,GoreePRL2008} statistics has been reported for both 2D and quasi-2D cases.

In this paper we show that a model that includes
only Yukawa interaction between particles, an externally imposed confining potential,  and stochastic heating and dissipation from neutrals, is able to demonstrate emergence of non-Gaussian statistical  features  associated with observed cooperative particle hopping  \cite{LinI1998,LinI2002,LinIPPCF2004,LinIPPCF2005, PoP} and vortical flow patterns \cite{PRL,NPJ}. The simulations also reveal some features for which we report the experimental footprint  here for the first time - the development of humps on the displacement probability distribution function (PDF) corresponding to integer multiples of the mean inter-grain distance  $\Delta$.

We study quasi-2D systems with varying ``stiffness", with particular attention paid to the transition in transport characteristics from time scales  $\tau\ll \tau_{\Delta}$ to $\tau>\tau_{\Delta}$, where $\tau_{\Delta}$ is the characteristic time for a dust grain to diffuse a  distance $\Delta$, i.e.  $\sigma(\tau_{\Delta})\equiv \langle \xi^2(\tau_{\Delta}) \rangle^{1/2}=\Delta$. For this purpose we analyze data from ground-based laboratory experiments (quasi-2D) and perform molecular dynamics simulations in quasi-2D and 2D  over a range of states, from pure crystal to liquid with particular focus on soft (viscoelastic or hexatic) states at the transition between crystal and liquid.

By  examining and contrasting these data, the following picture emerges: the shape of the position displacement distribution $P[\xi(\tau)]$ is Gaussian for times shorter than the grain-grain  collision time $\tau_c$. On these short time scales the grain motion is ballistic, and the displacement divided by $\tau$ can be interpreted as the instantaneous grain velocity. Hence the Gaussian displacement PDF reflects the Maxwellian velocity distribution resulting from the Gaussian stochastic heating term. On scales $\tau_c<\tau<\tau_{\Delta}$ the  PDF develops a leptokurtic shape. For stiffer systems, as $\tau$ approaches $\tau_{\Delta}$ humps develop on the tail of the PDF at displacements $\xi(\tau)$ corresponding to integer multiples of  $\Delta$. These humps signify that on these time scales the grain hopping in the imperfect hexagonal lattice influences the statistical transport characteristics. On even longer time scales, the kurtosis of the PDFs approach that of a Gaussian, indicating that multiple hoppings may represent a  sequence of independent transport events \cite{LinI2002}.

 In softer systems the picture on intermediate and  longer time scales is different. The humps due to hopping are masked by the emergence of large-scale vortical motions, and the PDFs on the hydrodynamical time-scales reflect these vortex motions.

An interesting observation is that Gaussian  statistics characterizes both the solidified crystalline state and the pure liquid state. Non-Gaussian statistics appears when the crystal is sufficiently soft to allow development of defects and particle hopping, but sufficiently stiff  to restrict  free diffusion on scales larger that $\Delta$. It appears on spatial displacement scales from a fraction of $\Delta$ and at least up to those scales where elasticity dominates over viscosity. The fact that Gaussian statistics and normal transport is restored in the purely liquid phase, indicates that the anomalous characteristics are associated with critical fluctuations in the form of a scale-free hierarchy of vortices  \cite{NPJ}. The vortices emerge from the stochastic forcing only if the lattice resides in a soft, viscoelastic transitional state between crystal and liquid.

We also discuss the possible physical origin of non-Gaussian displacement distributions and anomalous transport, and show how these feature derive naturally from a single-particle Langevin equation with a non-linear drift (friction) term modeling the collective effects of dust-dust interaction.

\section{Data sources}

\begin{figure}
\centering
\includegraphics[width=3.4in]{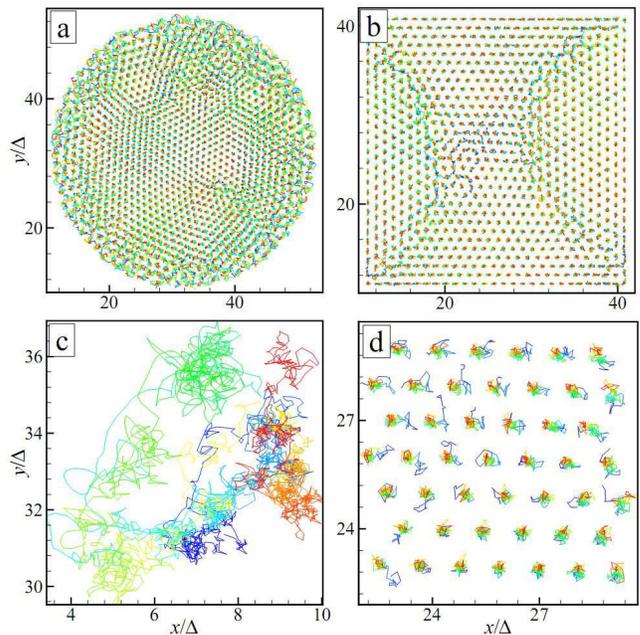}
\caption{(Color online) Grain trajectories from the molecular dynamics simulations of the Yukawa-Langevin system. The trajectories are color-coded by the time. Motion of particles during time $100 \, \delta t$ in the case of  parabolic confinement (a) and the hard wall confinement (b). A typical trajectory from the outer region of panel (a) plotted for time $4000 \,  \delta t$ (c). Close-up on stiff  central region of panel (b), motion of particles is shown for $100 \, \delta t$  (d).}
\label{sim_traj}
\end{figure}

\subsection{Laboratory experiments}
For this paper a re-analysis has been made of  raw data from the experiment reported in Ref.\  \cite{PoP}.  This experiment was performed in a capacitively coupled
rf discharge operated in argon at pressure 1.8 Pa and rf
power of 10 W. A cluster of 120 melamine-
formaldehyde monodispersive spheres with a diameter of
8.9 $ \mu $m, was confined by the potential created by a copper ring mounted on
the lower electrode. The dust particles were illuminated by a
horizontal laser sheet, and 50 000 images were taken by a
video camera at a sampling rate of 30 Hz and spatial resolution
of 6.7 $\mu$m/ pixel.

As seen from Fig. 1(b) of Ref.\ \cite{PoP}, where trajectories of particles over time lag of 1.7 sec are presented, the grains have limited mobility, and we refer to these data as our ``stiff cluster" experiment.
The re-analysis shows the  evolution of radial displacement PDFs  computed with
improved statistics compared to that of Ref.~\cite{PoP}. This new analysis reveals features of the PDFs which were not resolved in the earlier data, and also allows us to obtain PDFs of displacements on much longer time scales.

In Ref.\ \cite{PRL} a large circular cluster of 600 dust
grains in a monolayer configuration was studied in a similar experimental set-up. The cluster was brought to a partly ordered state  allowing large-scale viscoelastic vortex flows (see Figs.\  1 and  2 of Ref.\  \cite{PRL}). The experiment was
operated in argon at a pressure 4 Pa and rf power
of 19 W with melamine-formaldehyde spheres with diameter
of 7.2 $\mu$m. The grains were confined radially by the potential created by
a cavity of 6 cm radius machined into the lower electrode. The cluster diameter was 16 mm with interparticle
distance $\Delta\approx 0.6$ mm. In this experiment 30 000 images were taken by a video camera at  sampling
rate of 30 Hz and spatial resolution of 24 $\mu$m/pixel.  We will refer to this data as the ``soft cluster" experiment.

We will  also connect the results from our model simulations to the experimental studies on
 cold quasi-2D dusty-plasma liquids reported in Refs.\ \cite{LinI1998,LinI2002,LinIPPCF2004,LinIPPCF2005}.

\subsection{Langevin dynamics simulations}

The simulations are performed for the Yukawa system of particles embedded into a Langevin thermostat \cite{book} which simulates the effect of the collisions with the electrically neutral molecules of the background gas (details are given in \cite{MD}). The pair interaction between the dust particles is described by the Debye-H\"{u}ckel (Yukawa) potential: \be\phi(r) =
(Q/r) \exp(-r/\lambda_{\rm D}), \ee where $Q=e Z_d$ is the particle charge and $\lambda_{\rm D}$ is the the plasma screening length.
Particle radius, mass and charge, mean inter-particle distance, neutral gas density and temperature are the parameters of the simulations. Initially $N$ particles having the charge $Q$ are randomly distributed over a prescribed simulation volume to reach a value of (average) screening parameter $\kappa=\Delta/ \lambda_D$ close
 to that in the experiments.  After a transient phase, the system relaxes  to a statistically stationary state with the dust velocity distribution in thermal equilibrium with the Langevin thermostat. This state does not depend on the initial conditions, which is also the case in the experiments. Three-dimensional particle dynamics of the \textit{i}'th particle with mass $m$ evolves according to the equation,
\begin{equation}
m \frac{d^2 \textbf{r}_i}{dt^2}+ m \nu_n \frac{d \textbf{r}_i}{dt}=- Q\bigtriangledown{\phi}_c - Q\sum_{j\neq i} \bigtriangledown \phi(r_{ij})+\textbf{L}_i(t) \,\,,
\end{equation}
where  $\nu_n=1/\tau_n$ is the neutral gas friction, ${\phi}_c$ is the confinement potential and $r_{ij} \equiv \mid \textbf{r}_i-\textbf{r}_j \mid$. The stochastic Langevin force $\textbf{L}_i(t)$ (thermal noise induced by neutral gas particles) is defined from the relation:
\be \langle \textbf{L}_i(t) \cdot \textbf{L}_j(t+\tau) \rangle = 2 \nu_n m T \delta_{ij}\delta(\tau),\ee
with the mean zero condition $\langle \textbf{L}_i(t)\rangle=0$,
where $T$ is the neutral gas temperature.

Two types of the confinement have been used in the simulations; parabolic and hard wall.  For the parabolic trap the confinement potential is $\phi^{\rm p}_c(r) \propto r^2$. For the hard wall case the electric field associated with the confinement exponentially increases to the boundaries, so that $\phi^{\rm hw}_c(r) \propto \exp[(r - r_b )/\delta_{\rm w}]-1$ at $r>r_b$ and $\phi^{\rm hw}_c(r) = 0$ for $r < r_b$, where $r_b$ is the size of the system and $\delta_{\rm w}$ is a stiffness of the hard wall potential. In our simulations a value of $\delta_{\rm w} \simeq \Delta/3$ has been used.

To simulate a quasi-2D configuration  the motion in the vertical ($z$) direction has been suppressed  by a strong confinement in this direction. In most of the runs performed, the particle displacements were restricted to a range smaller than the mean horizontal interparticle distance $\Delta$, which is the typical situation for monolayer experiments. In addition, simulations of 2D systems (only horizontal $x$- and $y$- components considered) have also been run for both types of the confinement.

The parabolic confinement is in some respects a more realistic representation of  the experiments discussed here and, as seen from Fig.\ \ref{sim_traj}(a) it  reproduces flow fields  quite similar to those observed in the experiments. As in the experiment, such confinement results in a nonuniform cluster with more packed and stiff core and looser boundaries. The hard-wall potential, see Fig.\ \ref{sim_traj}(b), yields a different structure; particles are more packed at the boundaries (for more details see \cite{cnfkl1,cnfkl2,cnfkl3}). Ideally, comparison of these different confinement  schemes could help  identifying transport features which are robust to variations of the confinement.

As one can observe from Fig.\ \ref{sim_traj}(a) the resulting equilibrium state is a  cluster of dust grains in a partially ordered (hexatic) phase, similar to those reported from 2D  Yukawa MD simulations on square double-periodic domains in e.g. \cite{Reichardt2003} and in 2D hard-core simulations in \cite{Zangi}. However, our cluster is of finite size (non-periodic boundary conditions), confined by a circular symmetric parabolic potential, and is radially nonuniform. Like in the experiments the simulations are intended to describe, the transport over large distances takes place through development and propagation of 5-fold and 7-fold topological crystal defects.

By tuning the input parameters we obtain different states, ranging from pure crystal to pure liquid. As mentioned in the introduction we are mainly focused on semi-ordered (viscoelastic) states at the transition between crystal and liquid. To  obtain such states, the coupling parameter, defined as  $\Gamma_s=e^2 Z_d^2/(T \Delta) \exp(-\kappa)$, was changed in the range $50 - 100$ by varying $T$. The latter also implies change of the inter-particle distance; the corresponding range of the screening parameter was $\kappa\simeq 2-3$,  which is typical  for the experiments. For each state studied, $\Delta$ was determined as the distance to the first local maximum of the pair correlation function averaged over the cluster. This
distance varies less than $20\%$ from the core to the edge for
parabolical confinement, and is nearly uniformly distributed in space for hard wall confinement.

The particle positions have been sampled with a time resolution $\delta t$ chosen to be of the order of grain-grain collision time scale, $\delta t \simeq \tau_c$.  The grain-grain interaction time was estimated as $\tau_c\sim 1/(2 \pi \Omega_{E})$, where $ \Omega_{E}$ is the Einstein frequency which characterizes the linear oscillations of individual particles in a lattice and is a function of the screening parameter $\kappa$. For $\kappa \simeq 2 -3$ the Einstein frequency is about a factor of two of the dust-plasma frequency $\Omega_{0}=\sqrt{e^{2}Z^{2}/m \Delta^{3}}$ \cite{Knapek}.
The dust-neutral collision time scale was $ \tau_n \simeq 10 \delta t$ for all runs.

For each state, a simulation of $N=1000$ particles has been run during a time $\simeq 4000\,  \delta t$, thus proving adequate statistics for study of the system's evolution on long time scales.

In both experiments and  simulations the nature of the dynamics in the regimes where the system is in a semi-ordered state  is very sensitive to fine tuning of the control parameters, and we have not been able to reproduce all experimental features quantitatively. One reason for this may be that we are not able to fine tune parameters to the right state, but it can also be due to shortcomings of the model. However, the purpose of this paper is not to reproduce experimental results quantitatively, but to demonstrate that important qualitative features emerge from a simple Langevin model which does not involve other interactions between the dust particles and the plasma than the Debye screening of the Coulomb potential.

\section{Results on grain transport}

\subsection{Laboratory results}

\begin{figure}
\centering\includegraphics[width=3.25in]{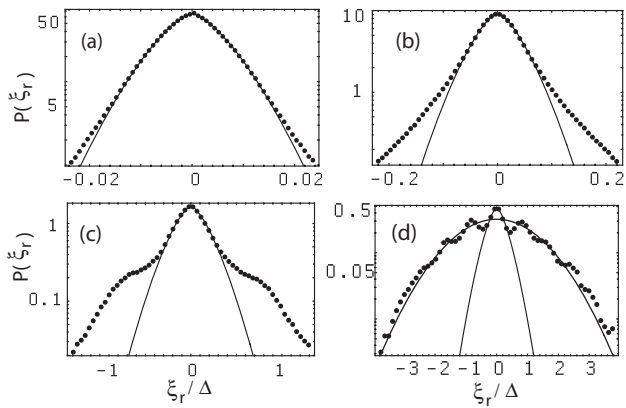}
\caption{Experimental results from stiff cluster (Ref.\cite{PoP}). PDFs of radial displacements $\xi_r(\tau) = r(t+\tau)- r(t)$ (dots) in units of
interparticle distance, $\Delta$, for different time lags; $\tau =0.033$~s (a), $\tau= 0.53$~s (b), $\tau= 8.5$~s (c) and $\tau=136$~s (d). The full curves are fits of the stretched Gaussian given by Eq.\ (\ref{pdf}) with $q=1.5$ (a-d), and the thick full curve is a Gaussian fit (d).}
\label{smallcluster}
\end{figure}

Time evolution of the PDFs of radial displacements $\xi_r(\tau) = r(t+\tau)- r(t)$ in units of
interparticle distance $\Delta$ for the stiff cluster is shown in Fig.\ \ref{smallcluster}.
The distributions are presented on a logarithmic scale and the deviation
from a Gaussian shape  on small time scales Fig.\ \ref{smallcluster} (a,b) is evident.
On those scales the shape of the PDF is well described by the stretched Gaussian
 \begin{equation}
P(\xi_r,\tau)=A(\tau) \exp[-B(\tau) |\xi_r |^p]
\label{pdf}
  \end{equation}
where $p=1$ corresponds to an exponential distribution and $p=2$ to the Gaussian. For the data presented the nonlinear fit (thin full lines) yields
the exponent $p\approx 1.5$.

 A possible, and rather trivial, explanation of the stretched gaussian PDFs would be the spatial non-uniformity of the system. For instance, the dust grains in the  laboratory clusters are more mobile close to the edge than in the central region. Thus, it is conceivable that a more heavy-tailed PDF could emerge from combining several Gaussian PDFs with different variances. We have tested this in the laboratory data by producing separate PDFs for particle displacements in the core and edge regions, respectively. However, although  these PDFs have different variances, they both retain essentially the same stretched-Gaussian PDF, and hence this is an intrinsic property of the particle dynamics on the short scales both in core an the edge, and is not an artifact due to spatial non-uniformity.

 When a significant number of particle displacements approach the mean   inter-grain distance $\Delta $ the PDF starts to develop ``shoulders" around $\Delta$ (Fig.\ \ref{smallcluster}(c)).
  The transition towards a Gaussian (Fig.\ \ref{smallcluster}(d), thick line) on the longest time scales involves a regime  where the PDF develops ``humps" at $\xi_r \approx n\Delta$, where $n$ is a positive integer.

 The humps on the PDFs are most easily explained as an effect of grain ``hopping" due to defects in the hexagonal crystal structure.
The humps can be observed only in relatively stiff systems where the hopping transport is not masked by the fluid vortical motion.
This is clearly illustrated if we compare results for the stiff cluster (Fig.\ \ref{smallcluster})  with those for
the softer cluster. As seen from Fig.\ \ 3 of Ref.\ \cite{PRL}, in the latter experiment, PDFs are well described by Eq.\ (\ref{pdf})
with $p=1.3$ on the small time scales, and a transition to Gaussian shape is observed already at the scales of $\xi_r \sim \Delta$.
On very large scales, $\sim 10 \Delta$, new humps emerge (Fig.\ 4(a) of Ref.\ \cite{PRL}) which
are due to large-scale vortical motion \cite{PRL,NPJ}.

In other experiments on quasi-2D laboratory dusty plasma similar results have been found. In studies of
 cold quasi-2D dusty-plasma liquids, Refs.~\cite{LinI1998,LinI2002,LinIPPCF2004,LinIPPCF2005} reported non-Gaussian
PDFs on scales below $\Delta$. Moreover, we find that the
central portion of the PDF presented in Fig.~3 of Ref.~\cite{LinIPPCF2004} can be fitted by Eq.~(\ref{pdf}) with $p=1.15$. On longer time
scales, when particle diffuse to distances  $\sim \Delta$, the PDFs in those experiments approach a Gaussian shape.

In the next subsection we show that the picture descried above is consistent with the results of simulations of the Langevin dynamics model.

\subsection{Langevin dynamics simulation results}

\begin{figure}
\centering
\includegraphics[width=3.25in, height=3.9in]{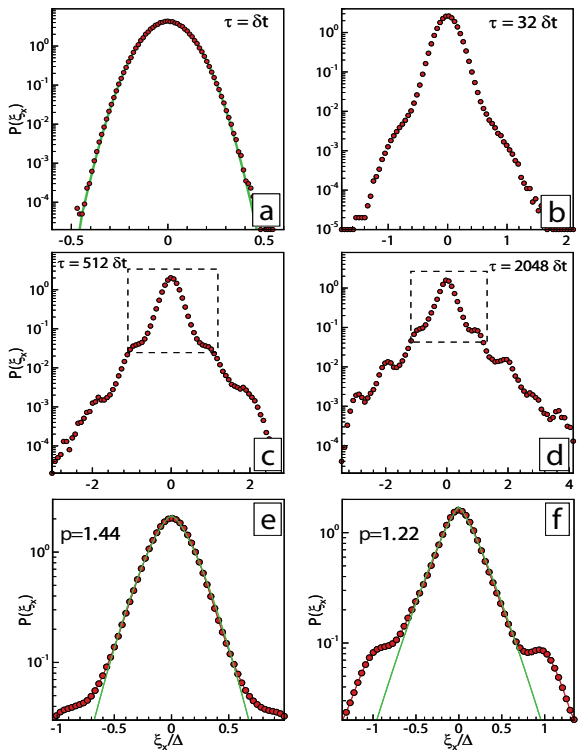}
\caption{(Color online) Langevin dynamics simulation of Yukawa system for the case of hard wall confinement. PDFs of horizontal displacements $\xi_x(\tau) = x(t+\tau)- x(t)$ (dots) in units of  $\Delta$, plotted for different time lags $\tau$ as indicated in the upper  corner. The full curve in (a) is a Gaussian fit and the full curves in the zoom-in figures (e,f) are fits by a stretched Gaussian given by Eq.\ (\ref{pdf}) with values of $p$ as indicated in the upper left corner.} \label{pdf_HW}
\end{figure}

The effect of the hexagonal lattice  structure is, as suggested above, most pronounced in the stiffer states. Langevin dynamics simulations with hard-wall confinement potential
reveals this quite clearly. In Fig. \ref{pdf_HW}(d) one observes  humps on the tail of the distribution  up to $\sim 4 \Delta$. Fig. \ref{pdf_HW}(a), shows the PDF of horizontal displacements on time scales $\delta t$, which is close to the collision time scale. On these time scales the grain motion is still ballistic, and the Gaussian PDF reflects the Maxwellian velocity distribution. On longer time scales the core of the distribution develops into a stretched Gaussian (see Fig. \ref{pdf_HW} (c,d) and zoom-in (e,f)), and shoulders (and later - humps) develop on the tail. These non-Gaussian features are obviously a result of hopping of some grains from one potential minimum in the crystal lattice to another.

It is interesting to compare these PDFs to those of the experiment displayed in Fig.\  \ref{smallcluster}.  The magnified core of the distribution in
Fig. \ref{pdf_HW}(d)  corresponds roughly to Fig. \ref{smallcluster}(c). In the simulations, however,  the  statistics is better because of the larger number of grains in the cluster, so humps on the far tail can be observed in  the zoomed-out image. These humps cannot be observed in Fig. \ref{smallcluster}(c) due to limited  statistics. They do appear, however at later times in the experiments, as shown in Fig. \ref{smallcluster}(c), when the width of the distribution exceeds $\Delta$. This figure has not been reproduced in the simulations so far, because the simulations would have to run considerably longer.

We have also analyzed the displacement PDFs for a sub-ensemble of particles in the core of the hard-wall cluster (see Fig.\ \ref{sim_traj}(d)). These particles are completely trapped in their potential wells in the lattice. It turns out that in such a purely crystalline state the displacement PDFs on all time scales are Gaussian.

\begin{figure}
\centering
\includegraphics[width=3.25in, height=2.7in]{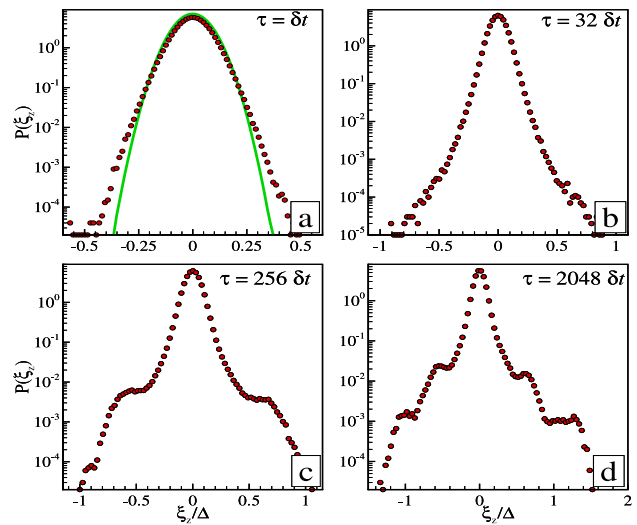}
\caption{(Color online) Langevin dynamics simulation of Yukawa  system for the parabolic confinement case (stiff cluster). PDFs of vertical $z$-component of displacements $\xi_z(\tau) = z(t+\tau)- z(t)$ (dots) in units of  $\Delta$, plotted for different time lags $\tau$ as indicated in the upper right corner.
 The  full curve in (a) is a Gaussian with the same standard deviation.}
 \label{pdf_PC_z}
\end{figure}

\begin{figure}
\centering
\includegraphics[width=3.25in,height=2.7in]{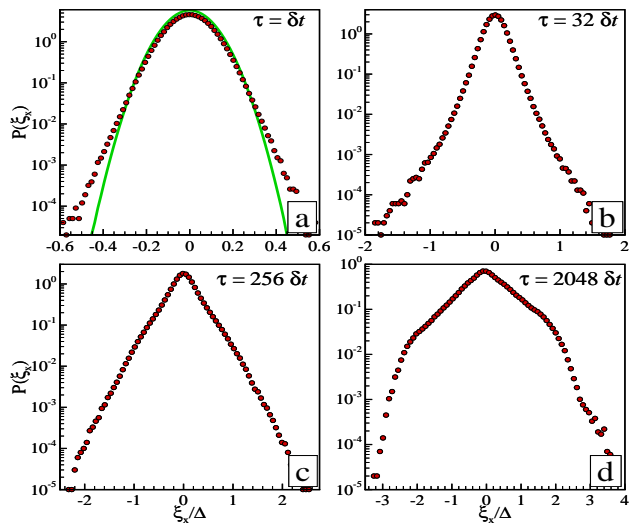}
\caption{(Color online) Same as in Fig.\ \ref{pdf_PC_z}, but now for the horizontal displacements $\xi_x$.}
\label{pdf_PC}
\end{figure}

Now let us discussed simulations performed for parabolic potential confinement. As discussed above, to simulate quasi-2D the motion in the vertical $z$ component is severely restricted. That results in the behavior of $z$-component of displacements similar to the hard-wall confinement case, as seen from Fig.\ \ref{pdf_PC_z}. The PDFs of horizontal displacements, shown in Fig.\ \ref{pdf_PC}, however still exhibit humps though much less pronounced because of the weaker confinement on $(x,y)$ plane. As shown in Fig.\  \ref{pdf_PC}(d) it approaches an overall Gaussian shape on longer time scales.

\begin{figure}
\centering
\includegraphics[width=3.25in,height=2.7in]{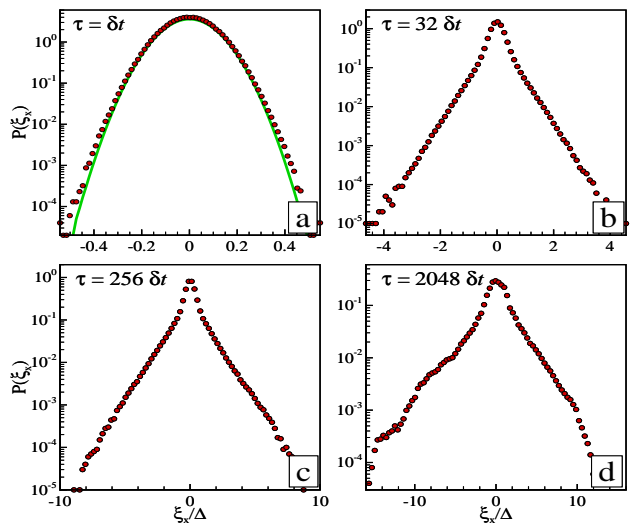}
\caption{(Color online) Same as in Fig.\ \ref{pdf_PC}, but now for the soft cluster.} \label{pdf_PC_soft}
\end{figure}

In simulations of even softer states, though still with preservation of the regular lattice structure on the smallest time scales, stretched Gaussian distributions still appear on the  time scales for which the standard deviation of the displacements is less than $\Delta$, but on longer time scales the PDF tends to develop humps on spatial  scales much larger than $\Delta$ (Fig.\ \ref{pdf_PC_soft} (d)). Such features were also observed in the soft experiment (see \cite{PRL}, Fig.\  4(a)), and was attributed to the formation of vortical flows on spatial  scales extending from $\Delta$ up to the cluster size.

This soft state has also been run in a pure 2D configuration, i.e. without $z$-component, with results similar to those presented in  Fig.\ \ref{pdf_PC_soft}. A preliminary fully 3D simulation of this state also yields qualitatively  similar results, and this indicates that dimensionality is not a critical issue. The qualitative features described here are robust and universal for Langevin dynamics of Yukawa clusters with partial preservation of crystalline order. These studies will be published in a  forthcoming paper.

Moreover, in these softer states, the system self-organizes into vortexes, or if stiffer, into rotating hexagonal domains which continuously break up and merge. The effect, called
``cooperative hopping",  was first reported in 2D dusty plasma clusters in \cite{LinI2002}. Large-scale vortexes driven by stochastic forcing have been observed experimentally \cite{PRL} and reproduced by the code used in the present study in Ref. \cite{NPJ}.
By plotting the particle positions and their pre-history over a suitable short time-interval can give an impression of the velocity-field of the vortical flow. If this is done in successive time frames one can construct movies of this velocity field,  which are very instructive to watch. Such movies for examples of experimental  and simulated flows can be found as auxiliary material to Ref.\ \cite{NPJ}. In Refs. \cite{PoP,PRL} one can find triangulation diagrams showing the occurrence of 5-fold and 7-fold defects associated with this hopping.

\section{Memory  effects and intermittency}

Information about the memory effects in the  transport characteristics can be obtained from  the variogram of particle displacements defined as
\begin{equation}
S(\tau)=(N-\tau/\delta t)^{-1}\sum_{j=1}^{N-\tau/\delta t}\xi_{x}^{(j)}(\tau)^2,
\label{var}
\end{equation}
where we have introduced the
position increments $\xi_{x}^{(j)}(\tau)=x(j\, \delta t+\tau)-x(j\, \delta t)$ over the time lag $\tau$.
 For  a fractional Brownian motion (fBm) the PDF of   $\xi_{x}^{(j)}(\tau)$ is Gaussian, and $S(\tau)= D\tau^{2 H}$, where $H$ is the Hurst exponent. Normal diffusion corresponds to Gaussian PDF and  $H=0.5$ (Brownian motion). $H$ different from 0.5 is usually referred to as sub- or superdiffusion, and indicates violation of the central limit theorem, which implies that either the displacements are not identical and independent random variables and/or each of them has a heavy-tailed distribution with infinite variance. The latter is not the case in our experiments and simulations. The limit  $H \rightarrow 1$
  corresponds to  ballistic motion.

In interpreting displacement variograms one has to realize that large-scale motions also influence the variogram on short time scales, so the Hurst exponent obtained on short-time scales does not necessarily reflect the nature of the diffusion of a particle relative to its nearest neighbors. Consider, for instance the extreme case of  a frozen ``flake'' of  dust grains which are completely fixed relative to each other, but that the flake itself is subject to transport characterized by a certain Hurst exponent. Then the variogram of every particle in the flake  is  characterized by this exponent, while their relative diffusion is negligible. In order to study this relative diffusion we would have to consider the evolution of relative position between neighboring particles.

Information about $H$ can also be obtained from the power spectral density of the particle displacement as a function of time. A power-law form of the spectrum, $S(f)\sim f^{-\beta}$, corresponds to an fBm with $H=(\beta-1)/2$. Thus normal diffusion requires  $\beta=2$.

The main departure from Gaussianity observed in our displacement PDFs is a flatter (more heavy-tailed) appearance, often well presented as a stretched Gaussian. A convenient measure of the flatness is the the kurtosis, which is the ratio between the fourth moment and the squared variance of the distribution. For the Gaussian the kurtosis is 3, and is higher for more heavy-tailed distributions. In the theory of turbulence, the concept of intermittency is used  for situations where the PDFs have higher kurtosis on small  than on large scales \cite{Frisch}. In our context intermittency would require that the kurtosis of  displacement PDFs decreases with increasing time scale.

\subsection{Laboratory results}
For the laboratory data of the stiff cluster, the power spectral density, autocorrelation function  and variogram of particle displacements are presented
in Figs.\ \ 3(a,b) of Ref.\ \cite{PoP}. In this experiment the PDF is non-Gaussian and self-similar on the short time scales,
 and the transport is superdiffusive with Hurst exponent $H=0.6$ (Fig.\  6(a) of Ref.\ \cite{PoP}). From results presented here, Fig.\ \ref{smallcluster}, it is apparent  that on scales $\tau\gg \tau_{\Delta}$ the displacement PDF approaches a Gaussian, so the kurtosis declines towards 3 at longer time scales. This indicates some intermittency - the displacements are more bursty on small scales than on large  scales.

For the softer cluster, the power spectra are presented in Figs.\ 4(c,d) of Ref.~\cite{PRL}  and  the variogram in Fig.\ 3(b). The transport is superdiffusive with $H=0.84$ on time scale $\tau< \tau_{\Delta}$ and with  $H=0.68$ for $\tau> \tau_{\Delta}$. By watching the movies of the particle motion one cannot avoid noticing that a common mode of motion the ``minimal vortex" consisting of a stationary dust grain about which its six nearest neighbors rotate. This constitutes a nearly ballistic motion on the time scale $\tau_{\Delta}$,  and might be the main contribution to the high value of  $H$ observed on scales $\tau\leq \tau_{\Delta}$. On longer time scales vortex motions on increasing scales are responsible for this memory-based superdiffusion. On the time scales one could follow the displacements in this experiment the PDF develops some large humps due to long-lived,  large vortices, and the tails are cut off by the finite size of the cluster. Hence, this experiment cannot tell us about the fate of the PDF on very large scale in extended systems.

\begin{figure}[h!]
\begin{center}
$\begin{array}{c@{\hspace{.1in}}c@{\hspace{.1in}}c}
\includegraphics[width=2.8in]{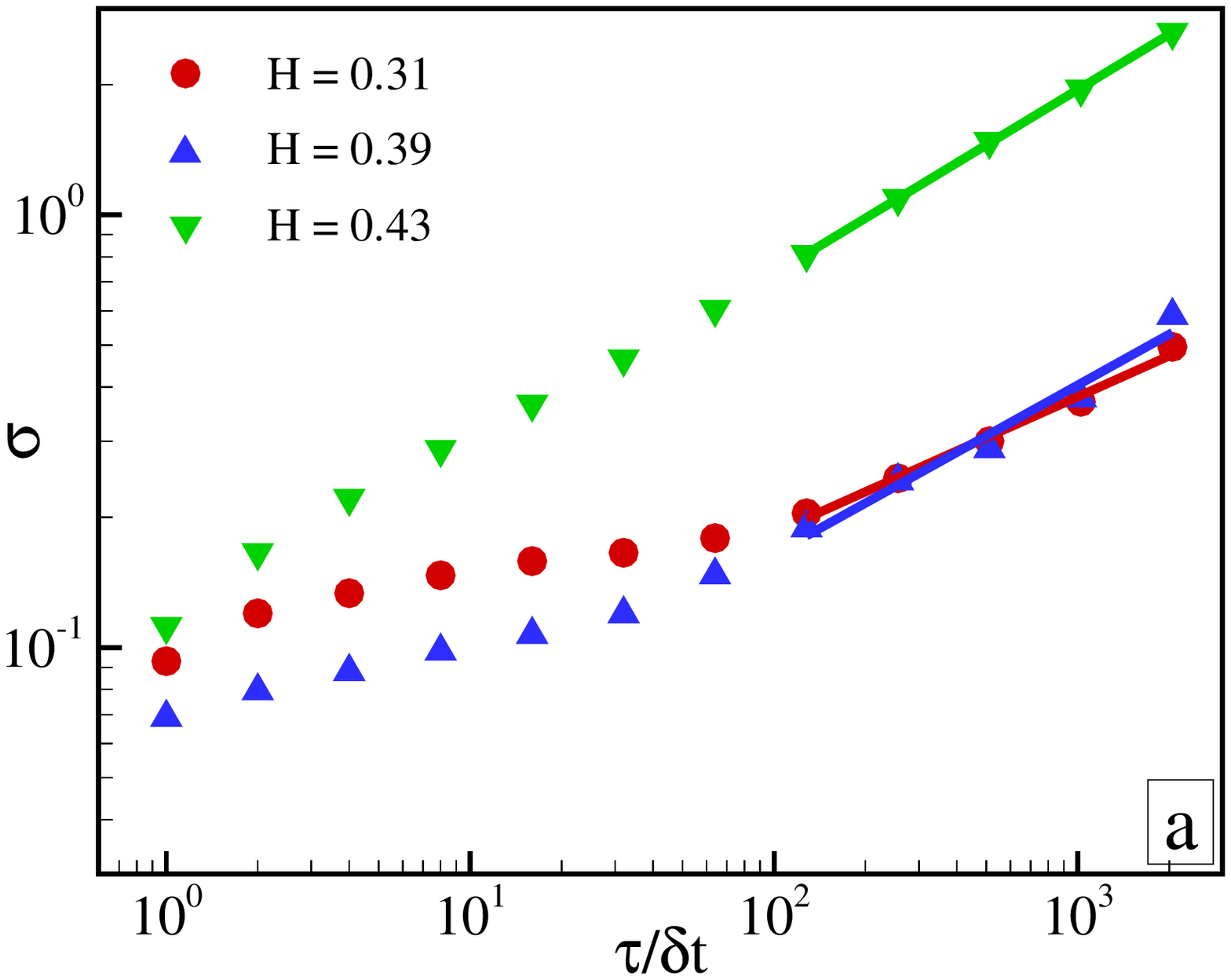}\\
\includegraphics[width=2.8in]{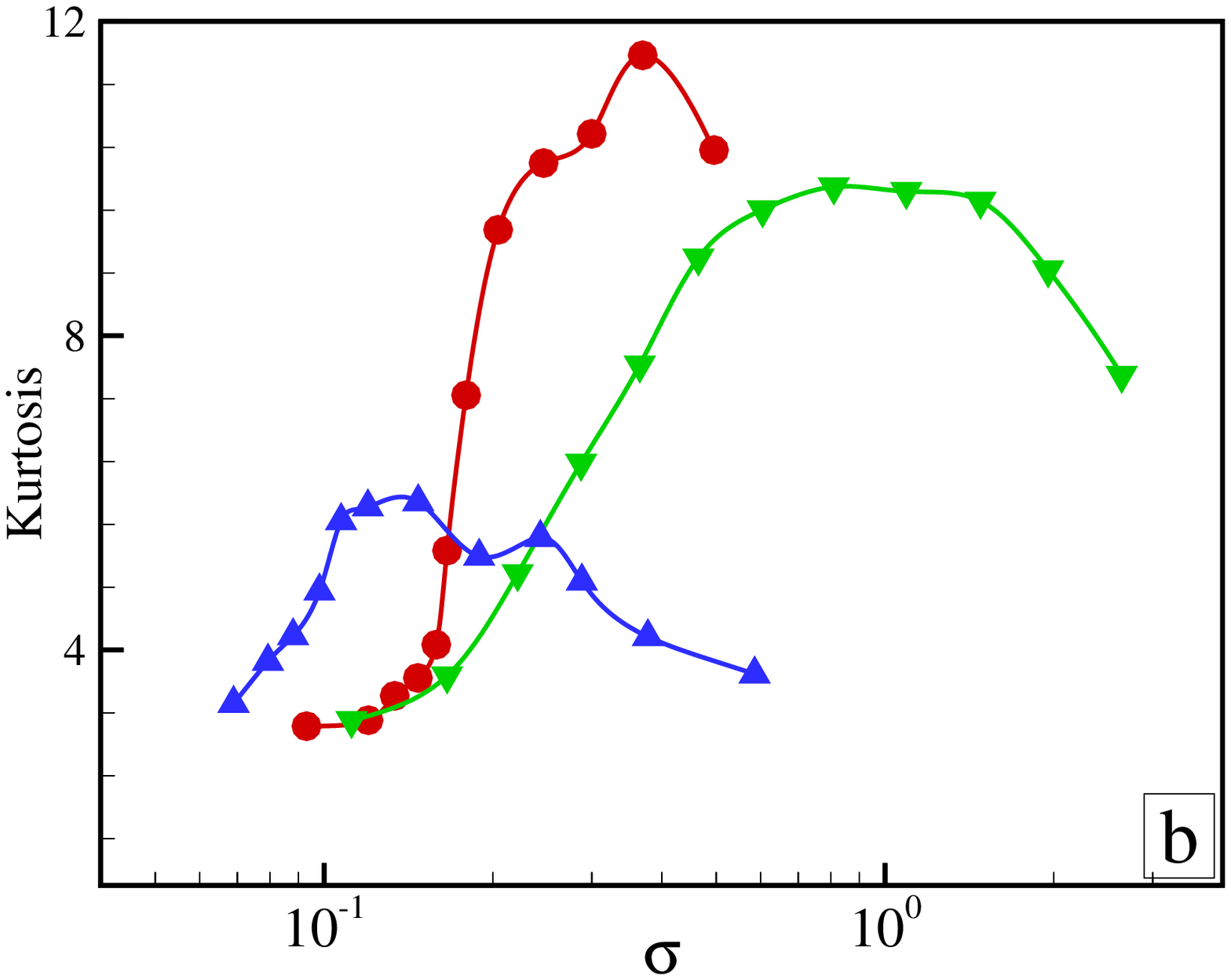}
\end{array}$
\end{center}
\caption{(Color online) Variograms $\sigma(\tau)=\sqrt{S(\tau)}$ in units of $\Delta$, where $S(\tau)$ is defined in Eq.\ (\ref{var}) (top panel (a)). ($\bullet$): Boundary region of hard-wall potential, corresponding to PDFs in Fig.\ \ref{pdf_HW}. ($\blacktriangle$): Parabolic confinement the case of stiff cluster,  corresponding to PDFs in Fig.\ \ref{pdf_PC}. ($\blacktriangledown$): Parabolic confinement, the case of soft cluster, corresponding to PDFs in Fig.\ \ref{pdf_PC_soft}. Kurtosis as function of $\sigma(\tau)$ (lower panel (b)). The symbols represent the same states as in panel (a).}
\label{vario}
\end{figure}

\begin{figure}
\centering
\includegraphics[width=2.8in]{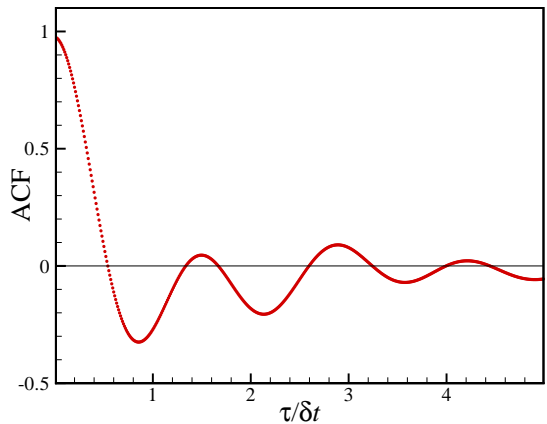}
\caption{(Color online) Displacement autocorrelation function (ACF) for the hard wall confinement case. The symbols represent the same state as in Fig.\ \ref{vario} (a). The sampling interval used is 0.01 $\delta t$.}
\label{AC}
\end{figure}
In the experiments on cold dusty-plasma liquids reported in Ref.~\cite{LinI2002} superdiffusive transport with $H\sim 0.65$ is found for  $\tau< \tau_{\Delta}$, and  normal diffusion ($H=0.5$) on larger scales.

\subsection{Langevin dynamics simulation results}

Variograms  for different runs with varying stiffness are presented in Fig.\ \ref{vario}(a). The variogram for the stiff core of hard-wall confinement  is flat after a certain time  ($H \rightarrow 0$), and on average particles do not travel more than $\sim 0.1 \Delta$, and is not plotted in the figure.

The curve represented by the $\bullet$ - symbols is representative for particles from the looser boundary regions of  hard-wall potential runs. Such particles are able to escape caging,  but the transport is subdiffusive ($H \rightarrow 0.3$) even for long time scales. The curve has an  ``S-shape" typical for cases when an oscillatory component is present in the time-series. The same should be reflected in autocorrelation function of the signal. Indeed, in Fig. \ref{AC} we find a clear periodicity on the relevant time scale.

For simulations of stiffer states with parabolic potential (the curve represented by ($\blacktriangle$) - symbols) the caging of particles is still noticeable for $\tau \leq 150 \, \delta t$ but for large times the Hurst exponent is higher,  $H \rightarrow  0.4$.

For simulations of soft states with parabolic potential the variogram (the curve with ($\blacktriangledown$) - symbols the effect of caging is less prominent, but in the limit of large $\tau$ we find also find  $H \rightarrow  0.4$ for this case.

The general picture for all states simulated (except when caging is total) is that  for $\tau<\tau_{\Delta}$ the PDFs are stretched Gaussians (with humps superposed for the stiffer states), and the diffusion is slightly subdiffusive - more so for the stiffer states. These distributions exhibit heavier tails than Gaussians, and hence kurtosis (flatness) higher than 3. For $\tau>\tau_{\Delta}$ the PDFs seem to tend more towards Gaussian again, at least when they are smoothed to remove the humps. This can be seen by plotting the kurtosis against the standard deviation
as done in Fig.\ \ref{vario}(b). It is observed that in general  the flatness increases with time as long as the standard deviation $\sigma(\tau)\equiv \langle \xi(\tau) \rangle^{1/2}\ll \Delta$, but have a clear descending trend towards the Gaussian value (3) as $\sigma(\tau)$ exceeds $\Delta$. For completely caged particles the kurtosis remains close to 3, confirming that the PDF remains Gaussian for these particles.

The  concept of intermittency in a signal is loosely defined as the tendency of a signal to be more bursty on short scales than on long scales \cite{Frisch}. This implies higher kurtosis on small than on large time scales. In our study this is the case on the ``hydrodynamic" scales, $\tau>\tau_{\Delta}$, but not on the scales of caging, $\tau\ll \tau_{\Delta}$.

\section{The origin of non-Gaussian displacements}
The Langevin thermostat used here models the interaction of the Yukawa many-body system with a heat bath of temperature $T$, and hence represents a canonical statistical  ensemble. Thus the simulated system should relax to a thermodynamic equilibrium. With proper tuning of control parameters our system develops strong fluctuations, which  in \cite{NPJ} was shown to exhibit (imperfect) scaling. This makes it natural to think about this as equilibrium critical behavior. Since this behavior is observed only when the control parameters are in a range where the system is subject to some local  positional order, we believe it is associated with a continuous solid-liquid phase transition. However, the small size and strong inhomogeneity of the systems simulated excludes direct comparison with theory \cite{review} and earlier simulations \cite{Hartmann} for such transistions in 2D Yukawa systems.

In extensive systems [{\em(systems without long-range interactions)}], thermodynamic equilibrium implies Gibbs-distributed  microstates of the cluster. This, however, does not exclude non-Gaussian distribution of particle displacements on varying time scales. In order to obtain some insight into the origin of this non-Gaussianity it may be useful to write a Langevin equation for a single dust grain in the form:
\begin{eqnarray}
\frac{d{\bf r}}{dt}&=&v, \\
\frac{d{\bf v}}{dt}&=&-\frac{Q}{m}[\nabla \phi ({\bf r})+\nabla \phi_{st}({\bf r},t)]-\nu_n {\bf v}+ \sqrt{D}\, \frac{d{\bf W}}{dt}, \label{Langevin1}
\end{eqnarray}
where   $\phi({\bf r})$ is the self-consistent (time-averaged)  caging potential well that a partially caged dust grain experiences, $\phi_{st}(\bf{r},t)$ is the additional fluctuating (stochastic) force field due to the Yukawa forces from the motion of the neighboring dust grains, $\nu_n$ is the dust-neutral   collision frequency, $D=2\nu T/m$ represents the diffusion coefficient due the stochastic force arising from collisions with neutrals, and ${\bf W}=(W_x,W_y,W_z)$ is the  vector of three independent Wiener processes (Brownian motions).

 Let us first recall some results on free diffusion of a dust grain imbedded in a neutral gas, i.e. let us  neglect the  term in square brackets on the right hand side of Eq.\ (\ref{Langevin1}). For a particle  starting at rest at the origin at time $t=0$, the equations can be integrated to yield  \cite{Gardiner} $\langle r^2(t) \rangle \rightarrow (2 T/m\nu_n )\, t$,
for $t\gg \tau_n\equiv \nu_n^{-1}$. On time scales $\tau\ll \tau_n$ the collision term $\nu_n {\bf v}$ is unimportant, and the velocity is a Wiener process. On these time scales the position process is the integral of the Wiener process, and is ballistic. The variogram would yield $H=1$.
For $\tau\gg \tau_n$ the velocity increments will be independent and Gaussian (white noise), and the position increments (displacements) will be Gaussian  and a Brownian motion, i.e. the slope is  $H=1/2$ in the variogram.

If we introduce a stationary potential structure, but  still neglect the stochastic force due to grain-grain collisions, there is a well-defined Fokker-Planck equation formulated in 6-dimensional phase space associated with Eq.\ (\ref{Langevin1}), which features the Gibbs distribution
$P({\bf r}, {\bf v})\sim \exp[-(\frac{1}{2}mv^2+Q\phi({\bf r}))/T]$ as a stationary solution, and in configuration space we have the Boltzmann-distribution
$p({\bf r})\equiv \int P({\bf r}, {\bf v})\, d^3v\sim \exp{[-Q\phi({\bf r})/T]}$.

If the caging potential barriers are much higher than the thermal energy of the grains, the cluster is in a completely frozen crystalline state, there is no diffusive  transport beyond the mean interparticle distance.

The omission of the stochastic force due to grain-grain collisions is only valid on time scales $\tau\ll\tau_c$, where $\tau_c$ is essentially the time it takes for a grain to bounce off the walls in its cage. The typical situation in our experiments and simulations is that $\delta t \sim \tau_c\ll\tau_n$. On time scales $\tau< \tau_c\ll\tau_n$,  as explained above, the motion is ballistic. If the velocities are Maxwellian  due to the Langevin thermostat, the displacements $\xi(\tau)\sim v\tau$ will be Gaussian distributed for $\tau<\tau_c$. This is what we observe in the simulations. In the experiments discussed in this paper we observe non-Gaussian displacements already for $\tau=\delta t$, but this is because the sampling time in the experiment is somewhat longer than the grain-grain collision time. This is supported by the results reported in \cite{Knapek}, where Gaussian displacements were found for $\tau=\delta t$ in an experiment where particle positions were sampled at 500 Hz.

On  time scales $\tau>\tau_c$ a single-particle stochastic equation makes sense only if the force due to grain-grain interaction is represented as a stochastic source term. If the issue under consideration is grain transport, the stochastic equation can be simplified by averaging the equations over a time scale of the order $\tau_n$, and obtain an equation for the averaged position. This justifies to neglect the inertial term $d{\bf v}/dt$ in Eq.\ (\ref{Langevin1}) (the Smoluchowski approximation), which then reduces to the
\begin{equation}
\frac{d{\bf r}}{dt}=-\frac{Q}{m\nu_n}\nabla \phi ({\bf r})+\frac{{\bf F}_{st}}{m\nu_n}, \label{Langevin2}
\end{equation}
where ${\bf F}_{st}$ represents the stochastic force due to collisions with neutrals as well as other dust grains.
A problem is of course that the interaction ${\bf F}_{st}$ contains long-range spatiotemporal dependencies and the statistical distribution of momentum transfers may be non-Gaussian, and hence cannot reliably be represented by a white noise term. One could certainly make assumptions about the statistical nature of the interaction, but the standard derivations of the associated Fokker-Planck equation are no longer valid since the position is no longer a Markov process.

Another approach would be to assume that the time-averaged potential $\phi$ is constant in the interior of the cluster, and to incorporate the the effects of the dust-dust interaction in the friction term and the stochastic term in Eq. (\ref{Langevin1}). The resulting stochastic equation would take the form (for simplicity we write the one-dimensional version):
\begin{equation}
\frac{dv}{dt}=K(v)+\sqrt{D(v)} \, \frac{dW}{dt}. \label{Langevin3}
\end{equation}
where the drift term $K(v)$ represents a generalized friction due to combined action of the dust-dust collisions and the dust-neutral collisions, and the second term on the right represents  the stochastic part of these interactions. In the following we shall assume that the latter contains no long-range memory and can be represented as Gaussian white noise.
An equation  of this form was studied by Kaniadakis and Quaraty \cite{ Kaniadakis}, and by Borland \cite{Borland} in the context of determining the conditions $K(v)$ and $D(v)$ must satisfy for the resulting stationary velocity PDF to reduce to the Tsallis distribution\cite{Tsallis1}. If the stochastic equation is interpreted in the sense of It\^{o} (see for instance \cite{Gardiner}), the associated Fokker-Planck equation for the velocity distribution $P(v,t)$ is
\begin{equation}
\frac{\partial P(v,t)}{\partial t}=-\frac{\partial\, }{\partial v}[K(x)P(v,t)]+\frac{1}{2}\frac{\partial^2\, }{\partial v^2}[D(v)P(v,t)]. \label{FP}
\end{equation}
If the drift term has a form that confines the particles to a range with finite velocities (which will be the case for a physically meaningful friction term) the stationary solution to Eq.\ (\ref{FP}) can be written as

\begin{equation}
P(v)=N\exp{\left( \int_{-\infty}^{v}\frac{2}{D(u)}\left[ K(u)-\frac{1}{2}\frac{\partial D(u) }{\partial u}\right] \, du\right)}, \label{sol-FP}
\end{equation}

where $N$ is a normalization constant. In the case of linear friction $K(v)=-\nu_c v$ and constant diffusion coefficient $D(v)=D$ Eqs.\ (\ref{Langevin3}) and (\ref{FP}) represents the well-known Ornstein-Uhlenbeck problem, and the solution (\ref{sol-FP}) is the resulting Maxwell velocity distribution. Borland \cite{Borland} shows that a general criterion for the solution (\ref{sol-FP}) to reduce to the Tsallis distribution \cite{Tsallis1}
\begin{equation}
P_q(v)=Z_q^{-1}[ 1-\beta(1-q)v^2]^{1/(1-q)], \label{Tsallisdistr}}
\end{equation}
is that $\{K(v),D(v)\}$ satisfy the relation
\begin{equation}
\frac{1}{D(v)}\left( K(v)-\frac{1}{2}\frac{\partial\, }{\partial v}\, D(v)\right) =-\frac{\beta v}{1-\beta(1-q)v^2} \label{Tsalliscond}
\end{equation}
Here $\beta=1/2kT$ is the inverse temperature and $q$ is the non-extensivity parameter. For $q=1$ the distribution reduces to the Maxwellian, and the family of functions that gives this conventional extensive Gibbs-Boltzmann statistics is given by the relation
\begin{equation}
K(v)=\frac{1}{2}\frac{dD}{dv}-\beta Dv. \label{Tsalliscond2}
\end{equation}
For constant diffusion coefficient this reduces to $K(v)=-\nu_c v$, where $\nu_c$ satisfies the relation $\nu_c=D/2kT$, which is a version of the fluctuation-dissipation theorem.

\begin{figure}
\centering
\includegraphics[width=3.0in]{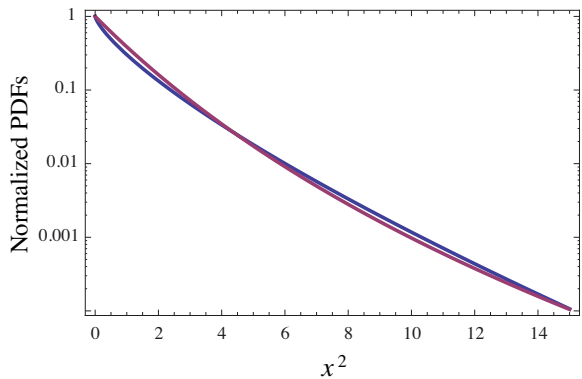}
\caption{(Color online) The red curve is a log-plot of the stretched-Gaussian function $\exp(-B x^{p})$ vs. $x^2$ for $B=1.2$ and $p=1.5$. The blue curve is a corresponding plot of the Tsallis distribution $[1+(q-1)x^2]^{-(q-1)^{-1}}$ for $q=1.1$. The values for $p$ and $q$ are typical for the experimental results discussed in this paper.}
\label{Tsallis}
\end{figure}

Borland emphasizes that Eq.~(\ref{Tsalliscond}) implies that there is a whole family of functions $\{K(v),D(v)\}$ which give rise to Tsallis statistics, but it should maybe also be mentioned that there exists an infinitely larger family of functions that do not. Thus, the ubiquity of Tsallis statistics in this context relies on the existence of ubiquitous physical situations where Eq.~(\ref{Tsalliscond}) is fulfilled. The analysis of our  experiments and simulations shows that the displacement PDFs are well  described by stretched Gaussians with $p\approx 1.5$ on time scales in the range $\tau_c<\tau<\tau_{\Delta}$, and  by Gaussians in the range $\tau>\tau_{\Delta}$. In Fig.\ \ref{Tsallis} we demonstrate that the stretched Gaussian with $p=1.5$ is practically indistinguishable from the Tsallis distribution with $q=1.1$, so there is no information in the experimental data to prefer one to the other. On the other hand the stretched Gaussian provides a very simple and physically interpretable generalization of the Ornstein-Uhlenbeck problem, as will be described in the following.

Let us consider the It\^{o}-Langevin equation Eq.\ (\ref{Langevin3}) with constant diffusion coefficient $D$, and generalized friction term\begin{equation}
K(v)=-\mbox{sgn}(v)\nu_c |v|^{p-1}, \label{genfriction}
\end{equation}
The physical rationale for   such a sub-linear friction term is simply to  account for  the reduced effective  dust-dust friction introduced by those fast particles undergoing cooperative hopping. It has been shown \cite{Reichardt2003,Zangi} that this hopping takes place along string-like structures of defects, and hence it is more natural to model it as a modification of the directed friction force than as modification of the diffusion coefficient.

In Eq.\ \ref{genfriction} the case $p=2$ corresponds to the standard Ornstein-Uhlenbeck problem, and $p=3/2$ would correspond to a friction that grows as the square root of the velocity, rather than proportional to the velocity. The stationary solution to the associated Fokker-Planck equation given by Eq.~(\ref{sol-FP}) reduces to the stretched Gaussian
\begin{equation}
P_p(v)=N_p\exp{[-(2\nu_c/p)|v|^p)] }. \label{strgauss}
\end{equation}
At this point we should keep in mind that the stretched Gaussian Eq.\ (\ref{pdf}) found in our data are PDFs for the position displacements at varying time lags $\tau$, not the velocity PDFs. In order to find the displacement PDFs we have integrated numerically the stochastic equation (\ref{Langevin3}) with the generalized friction term (and $p=1.5$) for an ensemble of initial velocites drawn from the stretched-exponential distribution Eq.\ (\ref{strgauss}). Then we have found the position trajectories by integrating $v(t)$, and formed the displacement PDFs for different time lags. The rather obvious result is that the displacement PDF remains close to the stretched Gaussian for $\tau\sim  \tau_c\equiv \nu_c^{-1}$ and converges to a Gaussian for $\tau \gg \tau_c$. The reason for this is that the displacements are near ballistic for $\tau< \tau_c$, and hence reflect the velocity distribution. On time scales greater than $\tau_c$ the velocity process is de-correlated and stationary, and successive displacements can be considered as independent, identical random variables.  Displacements $\xi_{\tau}$, for $\tau \gg \tau_c$, can therefore be considered as a sum of identical, independent random variables, and thus the conditions of the central-limit theorem are satisfied. Thus the displacement PDF should be Gaussian and the diffusion should be normal (Brownian motion).  The variograms produced on the basis of the computed PDFs shows (obviously) that $H=1$ for $\tau\ll \tau_c$ and $H=0.5$ for $\tau \gg \tau_c$, but also that there is a  large transitional regime  of time scales $\tau_c<\tau<50\tau_c$ where the slope corresponds to Hurst exponents in the range $0.5<H<1$.

This extremely simplified stochastic model reproduces the stretched exponentials observed for $\tau_c<\tau < \tau_{\Delta}$, and the convergence to Gaussian for $\tau\gg  \tau_{\Delta}$ observed in our simulations. Since the sampling time in the simulations are of the order of $\tau_c$ the variograms in Fig.\ \ref{vario} do not show the ballistic regime where $H=1$, but the behavior of the variograms in the regime $\tau \gg  \tau_{\Delta}$ can be interpreted as a slow convergence towards a Brownian motion ($H=0.5$). It is also noteworthy that the simulations in \cite{Zangi,Reichardt2007} gave rise to similar variograms  when the system was in the hexatic phase between a solid and a liquid. Those  simulations were performed with periodic boundary conditions, so the effects of the finite systems size and the system inhomogeneity  did not influence the variogram at large times as  they do in our simulations. Since the simulation in \cite{Zangi} also displays a leptokurtic distribution on the scales $\tau \leq \tau_c$, and a transition to a Gaussian for $\tau>\tau_c$, we conclude that the stochastic model, and the associated Fokker-Planck equation, yields a reasonable lowest-order description of anomalous diffusion in 2D many-body systems in the hexatic phase.

The linear Fokker-Planck approach discussed above (including the cases discussed by Borland) implies Gaussian displacement distribution in the limit $\tau \rightarrow \infty$. Hence it cannot be used to describe anomalous diffusion in the long-time limit. Another formulation that allows non-Gaussian displacement distributions and anomalous transport is the nonlinear Fokker-Planck equation \cite{Plastino,Tsallis2,Borland2},
 \begin{equation}
\frac{\partial P(x,t)}{\partial t}=-\frac{\partial\, }{\partial x}[K(x)P(x,t)]+\frac{D}{2}\frac{\partial^2\, }{\partial x^2}[P^{2-q}(x,t)]. \label{nl-FP}
\end{equation}
 In order to describe anomalous  displacement diffusion $x$ should now be interpreted as a spatial variable, and we should focus on time-dependent solutions. For this particular nonlinear diffusion term, and  a linearly varying drift term $K(x)=k_1+k_2x$, such a time-dependent solution takes the form of a Tsallis distribution, Eq.\  (\ref{Tsallisdistr}), where $\beta(t)$ and $Z_q(t)$ now are time dependent. In the case of free diffusion $k_1=k_2=0$, the variance $1/\beta(t)$ of the distribution  has the time-asymptotic dependence
 \begin{equation}
 \sigma_v^2\sim t^{\alpha}, \, \, \alpha=\frac{2}{3-q}, \label{Tsallisrelation}
 \end{equation}
 provided $q<5/3$. For $q>1$ the distribution has a power-law tail $P_q(x)\sim x^{2/(q-1)}$, and for $q\geq 5/3$ the tail is so heavy that the variance does not exist. For $1<q<5/3$ we have power-law tail and  $\alpha>1$, implying  superdiffusion. For $q=1$ the distribution reduces to a Gaussian and we have normal diffusion, and for $q<1$ the Tsallis distribution has a cut-off in the tail ($P_q(x)=0$  for $x>[1-\beta (t)(1-q)]^{-1/2}]$), and the transport is subdiffusive. Thus the Tsallis relation (\ref{Tsallisrelation}) implies that distributions with tails heavier than the Gaussian (leptokurtic) must be associated with superdiffusion.  Our simulations are subdiffusive for $\tau>\tau_{c}$, but the PDFs are leptokurtic and well described by the Tsallis distribution with $q\approx 1.1$, which according to Eq.\ (\ref{Tsallisrelation}) should give a weakly superdiffusive transport. This inconsistency indicates that the nonlinear Fokker-Planck equation is not a good description of the transport observed in our simulations.

There are also considerable difficulties associated with the physical interpretation of Eq.\ (\ref{nl-FP}) for our case. Borland \cite{Borland2}
 attempts to provide such an interpretation by considering $D\propto P^{1-q}$  as a $P$-dependent diffusion coefficient, that could replace $D$ in the stochastic equation (\ref{Langevin3}) (since we are considering spatial diffusion, the variable  $v$ in that equation should now be interpreted as a spatial variable, so we are really dealing with the Smoluchowski equation (\ref{Langevin2})). If $P(x,t)$ could be interpreted as a density, there would be no problem to envisage a density-dependent diffusion coefficient, but the density of dust grains in our system is almost constant in space and time, so a spatial diffusion equation for density  does not make sense. $P(x,t)$ will have to be interpreted as  a probablity density for  an ensemble of independent particle trajectories starting out  at positions selected  from  a prescribed initial distribution $P(x,0)$ in different  independent realizations of the experiment. Since we are interested in the evolution of particle displacements a natural choice is to consider an ensemble of particles starting from the same postition $x_0$, i.e.  $P(x,0)=\delta (x-x_0)$. The quantity $P(x,t)$ obviously depends on the choice of initial postion $x_0$, and  therefore $D\propto P^{1-q}(x,t)$ cannot be a local physical property of the medium at the point  at $(x,t)$. It is rather a quantity that is valid only for the particles belonging to this specific ensemble (i.e. the particles starting out at the position $x_0$ at $t=0$), and hence such a model would  imply that the diffusion properties of a particle depend very strongly on its prehistory, i.e. there must be strong long-range memory in the system.

\section{Conclusions}
In this paper we have studied anomalous transport properties; non-Gaussian particle displacement statistics and memory effects, in a Langevin-dynamics model, and we have  contrasted these  simulations with experimental results from quasi-2D dust plasma clusters in soft states. We have demonstrated that anomalous transport is ubiquitous on time scales $\tau_c<\tau \leq \tau_{\Delta}$ provided the system is not residing in the pure crystalline or liquid phases. Moreover, we have shown that this complex dynamics emerges from stochastic forcing and dissipation provided by the collisions with the neutral gas, and that no other plasma effects than the static Debye screening are  required.

From the experiments and the Langevin  dynamics simulations we observe that Gaussian displacement distributions are restricted to the  following cases:
(i) on time scales less than the collision time $\tau_c$ in all systems,
(ii) on all time scales for systems that are so stiff that no hopping takes place (pure crystal),
(iii) on all time scales for systems where no caging occurs (pure liquid or gas),
(iiii) for time scales longer than $\tau_{\Delta}$ in soft systems.

A result that seems to be robust for both stiff and soft states, is that PDFs evolve from a leptokurtic
(heavy-tailed) shape for small time-lags $\tau\ll \tau_{\Delta}$ towards a Gaussian shape for larger $\tau$. This feature is usually interpreted as a signature of
 intermittency - a spikyness in the flow-field which gives rise to more leptokurtic PDFs of the displacements on the small scales. Note that  intermittency does not prevail  to spatial scales less than $\Delta$. A simple way to model this intermittency feature is via the single-partice stochastic equation (\ref{Langevin3}), and the associated linear Fokker-Planck equation (\ref{FP}), with a non-linear drift term given by Eq.\ (\ref{genfriction}).

In Ref. ~\cite{PRL} it is pointed out that the dust monolayer should be perceived as a viscoelastic medium, where the elastic properties
 dominate on the smaller scales. Thus the transition from stretched-Gaussian to Gaussian PDFs could correspond to the transition from  elastic
 to viscous flow. One future goal would be to perform experiments and simulations on sufficiently large spatial and temporal scales to characterize the transport associated with homogeneous, turbulent flow in such systems. With such extensive simulations one could clarify, for instance, if energy will cascade to vortices on infinitely large scales in an infinite system, or if there is an upper limit to the vortex size related to the characteristic length over which elasticity dominates over viscosity.

 The rather imprecise qualifier ``soft" has been employed in this paper to characterize 2D dusty plasma clusters which preserve local hexagonal order, but exhibit loss of such order on larger  scales. Materials displaying such properties are also classified as ``hexatic", and the continuous transition  from the crystalline to the liquid state  might be considered as a continuous phase transition \cite{Christensen}. In \cite{NPJ} we perform a particular  avalanche analysis of the experimental soft cluster and a Langevin-dymamics simulation. In this analysis we define connected space-time regions where the kinetic energy of the grains exceeds a prescribed threshold, and define the  activity in these regions as avalanches. The statistics of avalanche sizes and durations tend to obey power laws with similar characteristic exponents for experiment and simulation, which are properties characteristic for critical phenomena and higher order phase transitions \cite{Christensen}. Such phenomena have been studied in strongly idealized models  such as the Ising model \cite{Christensen,Gitterman} and the XY model \cite{Gitterman}, but the Yukawa-Langevin model studied in this paper exhibits, in spite of its simplicity with respect to the physical interactions,  the prospect of quantitative prediction of dynamics in real physical systems where the microscopic state of the system can be tracked to the smallest  detail of physical interest.

\begin{acknowledgements}
S.R. acknowledges support of the Swedish Research Council. K. R. acknowledges support of the Norwegian Research Council under grant 171076/V30, and is grateful to Dr. Martin  Rypdal for helpful discussions and for providing numerical solutions to Eq. (\ref{Langevin3}). The authors also acknowledge several helpful comments from anonymous referees.
\end{acknowledgements}


\begin{thebibliography}{99}
\bibitem{review} V.E. Fortov  {\it et al},  Phys. Rep., {\bf 421}, 1 (2005).
\bibitem{Christensen} K. Christensen  and N. R. Moloney, {\em Complexity and Criticality}, (Imperial College Press, London, 2005).
\bibitem{Reichardt2007}C. Reichhardt and C. J. Olson Reichhardt, Phys. Rev. E {\bf 75}, 051407 (2007).
\bibitem{GoreePRE2008} B. Liu, J. Goree and Y. Feng, Phys. Rev E {\bf 78}, 046403 (2008).
\bibitem{Reichardt2003}C. Reichhardt and C. J. Olson Reichhardt, Phys. Rev. Lett{\bf 90}, 095504 (2003).
\bibitem{Zangi} R. Zangi and S. Rice, Phys. Rev. Lett {\bf 92}, 035502 (2004).
\bibitem{LinI1998} W.-T. Juan and L. I, Phys. Rev. Lett. {\bf 80}, 3073 (1998).
\bibitem{LinI2001} W.-T. Juan, M.-H. Chen, and L. I, Phys. Rev. E {\bf 64}, 016402 (2001).
\bibitem{LinI2002} Y.-J. Lai and L. I, Phys. Rev. Lett. {\bf 89}, 155002 (2002).
\bibitem{LinI2004} W.-Y. Woon and L. I, Phys. Rev. Lett. {\bf 92}, 065003 (2004).
\bibitem{LinIPPCF2004} Ying-Ju Lai, Wei-Yen Woo, and Lin I, Plasma Phys. Controlled Fusion {\bf 46}, B449 (2004).
\bibitem{LinIPPCF2005} Chia-Ling Chan et al., Plasma Phys. Controlled Fusion {\bf 47}, A273 (2005).
\bibitem{PoP} S. Ratynskaia, C. Knapek, K. Rypdal {\it et al.}, Phys. Plasmas {\bf 12}, 022302 (2005).
\bibitem{PRL} S. Ratynskaia, K. Rypdal, C. Knapek {\it et al.}, Phys. Rev. Let., {\bf 96}, 105010 (2006).
\bibitem{GoreePRE2007} Bin Liu and J. Goree, Phys. Rev. E., {\bf 75}, 016405 (2007).
\bibitem{GoreePRL2008} Bin Liu and J. Goree, Phys. Rev. Let., {\bf 100}, 055003 (2008).
\bibitem{NPJ} K. Rypdal, B. Kozelov, S. Ratynskaia, B. Klumov, C. Knapek, M. Rypdal, New Journal of Phys. {\bf 10} 093018 (2008).
\bibitem{Knapek} C. Knapek, A. Ivlev, B. Klumov, G. Morfill, and D. Samsonov, Phys. Rev. Let., {\bf 98}, 015001 (2007).
\bibitem{book}D. Frenkel, B. Smit, {\em Understanding Molecular Simulation, Academic Press}, (New York, 2002).
\bibitem{MD} B. A. Klumov, M. Rubin-Zuzic, G. E. Morfill, JETP Letters, {\bf 84}, 10, 542, (2007).
\bibitem{cnfkl1} B. A. Klumov, G. E. Morfill, JETP Letters, {\bf 85}, 10, 498, (2007).
\bibitem{cnfkl2} B. A. Klumov, G. E. Morfill, JETP Letters, {\bf 87}, 8, 409, (2008).
\bibitem{cnfkl3} B. A. Klumov, G. E. Morfill, JETP, 107, 5, 908 (2008)
\bibitem{Frisch} U. Frisch {\em Turbulence. The Legacy of A. N. Kolmogorov} (Cambridge University Press, Cambridge, 1995).
\bibitem{Hartmann} P. Hartmann, Z. Donk\'{o}, P. M. Bakshi, G. J. Kalman and S. Kyrkos, IEEE Trans. Plasma Science {\bf 35}, 332, (2007).
\bibitem{Gardiner} C. W. Gardiner {\em Handbook of Stochastic Methods}, (Springer, Berlin, 1983).
\bibitem{Kaniadakis} G. Kaniadakis and P. Quarati, Physica A {\bf 237}, 229 (1997).
\bibitem{Borland} L. Borland, Phys. Lett. A {\bf 245}, 67 (1998).
\bibitem{Tsallis1} C. Tsallis, J. Stat. Phys. {\bf 52} 479 (1988).

\bibitem{Plastino}A. R. Plastino and A. Plastino, Physica A {222},347 (1995).
\bibitem{Borland2}L. Borland, Phys. Rev. E, {\bf 57}, 6634 (1998).
\bibitem{Tsallis2} C. Tsallis and D. J. Bukman, Phys. Rev. E  {\bf 54}, R2197 (1996).
\bibitem{Gitterman} M. Gitterman and V. Halpern, {\em Phase Transitions. A brief Account with Modern Applications} (World Scientific, Singapore, 2004).
\end{thebibliography}
\end{document}